\documentclass[final,12pt,twocolumn]{elsarticle}
\usepackage[utf8]{inputenc}
\usepackage{textcomp}
\DeclareUnicodeCharacter{207A}{\textsuperscript{+}}

\usepackage{amssymb}
\usepackage{amsmath}
 \usepackage{xcolor}
\usepackage{gensymb}
\usepackage{hyperref}
\usepackage{url}
\usepackage{xurl} 

\journal{Journal of The American Ceramic Society}

\begin{document}
\begin{frontmatter}



\title{Upcycling solar glass into Ce‑doped oxyfluorides: spectroscopic and crystallization properties} 


\author[UTFPR,UEM,CEMHTI,studium]{Marcos Paulo Belançon\corref{cor1}}
\ead{marcosbelancon@utfpr.edu.br}
\cortext[cor1]{Corresponding author}
\author[UTFPR]{Rafaela Valcarenghi}
\author[UTFPR]{Marcelo Sandrini} 
\author[UEM]{Brenno Silva Greatti}
\author[UEM]{Robson Ferrari Muniz}
\author[UEM]{Vitor Santaella Zanuto}
\author[CEMHTI]{Sandra Ory}
\author[CEMHTI]{Aurélien Canizares}
\author[CEMHTI]{Maxence Vigier}
\author[CEMHTI]{Emmanuel Veron}
\author[CEMHTI]{Mathieu Allix}
\author[CEMHTI]{Michael Pitcher}

\affiliation[UTFPR]{organization={Universidade Tecnológica Federal do Paraná},
            addressline={Via do Conhecimento Km 01}, 
            city={Pato Branco},
            postcode={85503-390}, 
            state={Paraná},
            country={Brasil}}

\affiliation[UEM]{organization={Universidade Estadual de Maringá},
            city={Maringá},
            state={Paraná},
            country={Brasil}}

\affiliation[CEMHTI]{organization={Laboratoire CEMHTI, UPR 3079-CNRS},
            city={Orléans},
            country={France}}

\affiliation[studium]{LE STUDIUM - LE STUDIUM Loire Valley Institute for Advanced Studies}

\begin{abstract}
Oxyfluorides made with up to 80~wt\% of recycled glass from end-of-life solar panels have been investigated. Reduced processing temperature and high transparency have shown that the material has potential for optical applications. In this work, cerium-doped samples were investigated. Spectroscopic study reveals the presence of Ce$^{3+}$, and luminescence from these ions and oxygen-deficient centers was detected. Raman demonstrated that cerium affects the glass network by promoting polymerization. In turn, thermal analysis indicated some changes in the crystallization events between 500-800~\degree C, which were confirmed by in situ X-ray powder diffraction measurements. Crystallization of fluorite, xonotlite, and combeite was confirmed, while other phases give minor contributions to the XRD patterns. Cerium addition reduced the formation of xonotlite, mainly above 700~\degree C. The potential applications of the material and the further studies required are discussed.
\end{abstract}


\begin{highlights}
\item End-of-life solar glass was used to produce cerium-doped oxyfluorides and Ce$^{3+}$ presence was confirmed by spectroscopic measurements
\item DSC, Raman, and in situ PXRD experiments confirmed that Cerium affected the material structure and crystallization dynamics
\item Fluorite, xonotlite, and combeite crystal phases were detected on a high-temperature PXRD study
\item The potential of this material for different applications is discussed
\end{highlights}


\end{frontmatter}



\section{Introduction}

The world has been experiencing rapid growth in solar electricity production, but researchers have shown that this rate is insufficient to meet our greenhouse gas reduction targets~\cite{Haegel2023}. The industry faces several challenges in scaling up the production of photovoltaic solar panels, including the production of flat glass itself~\cite{Chowdhury2025}.

More specifically, the flat glass used in solar panels, which from now on we will call ``solar glass'', requires high-purity raw materials compared to other applications, as iron and other impurities can compromise the transparency of the glass and, consequently, the electricity output of solar panels~\cite{FerrariMuniz2025,Belancon2023}.

Today's solar glass production is about 24 million tons annually, and it should be multiplied by 3-4 to keep pace with the expansion of photovoltaics and keep global warming below 2~\degree C~\cite{Chowdhury2025}. To put that in perspective, the total glass output of the European glass industry is about 40 million tons~\cite{Jost2025}. It is challenging to separate the glass from the other materials in end-of-life solar panels~\cite{Chowdhury2026}, and unwanted mixing impurities can inhibit the recycling of these materials~\cite{Inano2025,Li2024}. Other specific challenges are related, for example, to antimony, which has been used to produce solar glass~\cite{Chowdhury2026b} and may pose an environmental threat to some reuse routes for this material~\cite{Sandhu2013}.

To achieve more sustainable practices for end-of-life solar glasses, one potential alternative is to develop new glasses and glass-ceramics that use solar glass as a raw material. In a previous work, we reported the production of a series of oxyfluoride glasses (CgCAF), composed of up to 80~wt\% of cullet from solar glass~\cite{Valcarenghi2025}, which was previously extracted from a photovoltaic panel~\cite{francisnara2022}. The melting temperature of the samples decreased significantly upon the introduction of CaF$_2$ and Na$_2$CO$_3$, yielding high transmittance in a wide range which enables these samples to be explored for a variety of applications.

However, in some cases, we need to protect materials and devices from UV radiation, and thus the improved UV transparency of these oxyfluorides needs to be controlled. In this work, we aim to accomplish this by incorporating cerium into one of these oxyfluoride compositions and investigating whether it affects the material's optical and thermal properties and crystallization dynamics. In cerium-doped glasses, this element can assume two distinct valence states, Ce$^{3+}$ or Ce$^{4+}$~\cite{Milewska2025}. While the latter can induce strong UV absorption~\cite{Taniguchi2020,Sandrini2023}, the former exhibits a parity-allowed 4f-5d transition, which can result in intense luminescence. Often, though, a coexistence of both valence states can be challenging due to the overlap of optical absorption of those ions~\cite{Volotinen2022}, in such a way that Ce$^{4+}$ can even prevent Ce$^{3+}$ luminescence. 

As we are going to show next, our study on the incorporation of cerium oxide into the CgCAF matrix confirmed the presence of Ce$^{3+}$, and demonstrates that cerium oxide affects structural and crystallization properties. These findings are a fundamental step toward probing the properties of this glass at temperatures near and above the glass transition, as this may affect the material's processing. On the other hand, this study also provides an evaluation of the potential of this glass system to develop new glass-ceramics, which could be tailored for several applications, potentially enabling an upcycling route to solar glass from end-of-life photovoltaic panels.

\section{Materials and methods}

First, we selected the glass matrix system, previously referred to as CgCAF12~\cite{Valcarenghi2025}, due to its thermal stability against crystallization, which enables the production of samples in various shapes and sizes. This glass was obtained by mixing CaF$_2$ and Na$_2$CO$_3$ (Sigma-Aldrich
with 99.0~\% purity) with the solar glass and melting it at 1200~\degree C. The conventional melt-quenching process in air was applied, and the material was cast in a stainless steel mold preheated at 480~\degree C (below CgCAF12's T$_g$), where it was kept for 12 h to relieve internal stresses. This resulted in cylindrical samples measuring approximately 3~cm by 1~cm in height and diameter, respectively. Slices of the samples were cut and polished or ground for each experiment. More details about the process can be found in our previous work~\cite{Valcarenghi2025}.

The CgCAF12 matrix composition, confirmed by XRF measurements~\cite{Valcarenghi2025}, is about 58.2 mol\% SiO$_2$, 16.1 mol\% Na$_2$O, 12.4 mol\% CaF$_2$, 7.6 mol\% CaO, 3.6 mol\% MgO, 0.3 mol\% K$_2$O, and 0.3 mol\% Al$_2$O$_3$. Besides this, small amounts of SrO, ZnO, and P$_2$O$_5$ ($\sim$0.1 mol\%), and trace quantities ($<$0.05 mol\%) of Fe$_2$O$_3$, Cr$_2$O$_3$, MnO, PbO, BaO, TiO$_2$, and ZrO$_2$ were also detected, though, they are provenient from the solar glass and we found no increase in these impurities in the CgCAF12 matrix. Doped samples containing X~mol\% of cerium oxide were produced by the same process, resulting in a total of (100+X)~mol\%. These samples were named CgCAF12CeX, where X=0.10, 0.25, 0.50, or 1.00.

A JASCO FP-8550 fluorometer was employed to perform spectroscopic measurements in transparent disc samples. An integrating sphere (ISF-134) was used to obtain the absorbance from reflectance data. Emission and excitation measurements were performed using the standard cell holder (SCE-146), in which glass disc samples were mounted on a metallic support. The intensities of the spectra shown here cannot be compared across different samples and experimental conditions, as they were manually adjusted to facilitate data visualization and interpretation.

A Renishaw INVIA REFLEX Raman spectrometer with a 1800 lines/mm diffraction grating was used to obtain Raman spectra with a 514 nm laser. The disc glass samples were analyzed over a wavelength range of 200-1600 cm$^{-1}$.

DSC was performed in a SETARAM Multi HTC 1600. Powdered samples were placed in a Pt-Rh crucible and heated up to 850~\degree C with a heating rate of 10~\degree C/min under argon (20 ml/min).

In situ HT XRPD analysis was performed on powdered material to evaluate the temperature-dependent structural evolution. The measurements were performed in the Bragg–Brentano $\theta$-$\theta$ geometry on a high-flux SmartLab Rigaku rotated Cu anode ($\lambda$ = 1.5418\AA) diffractometer equipped with the HypiX-3000 HE detector and an Anton Paar oven chamber (model HTK 1200N, Graz, Austria), which allows measurements up to 1200~\degree C. Powders obtained from crushed glass beads were placed in a platinum-lined corundum sample holder. Data were collected from 500~\degree C to 800~\degree C with a 20~\degree C step size, an angular range of 10-80~\degree (2$\theta$) with a 0.016~\degree~step size, and a speed of 1~\degree/min. Next, we report the main findings of our investigation.

\section{Results and discussion}
\begin{figure}[htp]
    \centering
    \includegraphics[width=1\linewidth]{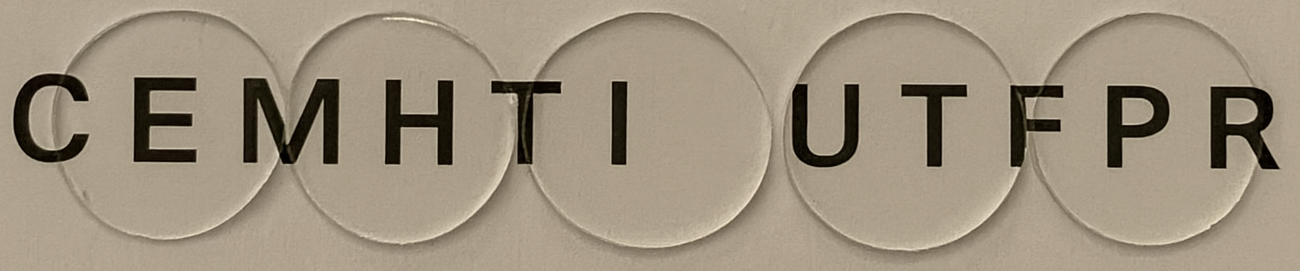}
    \includegraphics[width=1\linewidth]{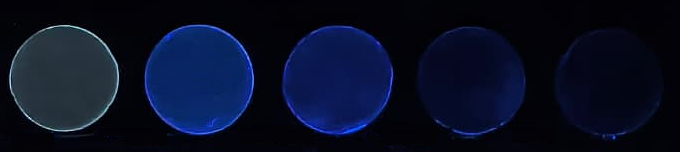}
    \caption{CgCaF12 samples under natural light (top) and 254~nm ilumination (bottom), from 0 to 1.0~\% (left to right) cerium oxide concentration.}
    \label{fig:placeholder}
\end{figure}

Figure~\ref{fig:placeholder} shows a photo of the CgCaF12 matrix and cerium-doped samples under 254~nm illumination.
The CgCaF12 glass exhibits a pale green emission under this UV excitation, while in the cerium-doped samples, the emission is predominantly blue. Silicates in general exhibit a mix of defects that affect UV optical absorption, which may still be linked to visible luminescence. The defects can be unpaired electrons called E-centers ($\equiv$Si$\bullet$), nonbridging-oxygen hole centers ($\equiv$Si--O\textbullet), peroxy radicals ($\equiv$Si--O--O\textbullet), self trapped electrons, oxygen-deficient centers ($\equiv$Si--Si$\equiv$) and others~\cite{Salh2011,Griscom2011}. In this way, even in the case of pure silica, we may have a variety of these defects~\cite{Leone2003,Zatsepin2010,Jackson2025}. 

\begin{figure}[htp]
    \centering
    \includegraphics[width=1\linewidth]{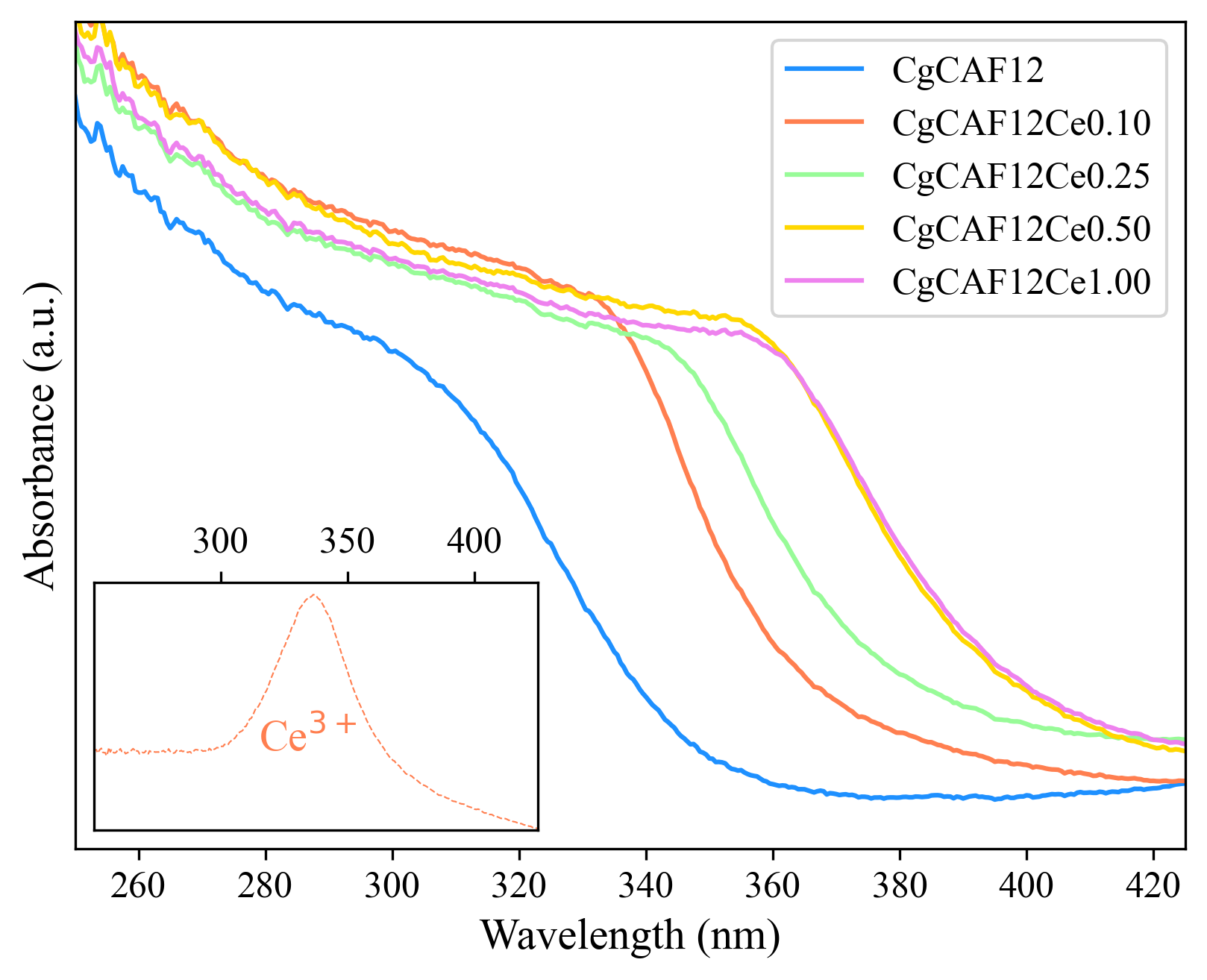}
    \caption{Absorbance spectra as measured. The inset shows the computed difference between CgCAF12Ce0.10 and CgCAF12 spectra.}
    \label{fig:abs}
\end{figure}

Concerning the more complex CgCAF12 composition, which also contains fluorine, it is even harder to fully identify the absorbing species. To investigate this feature, we obtained UV/Blue absorbance spectra from reflectance measurements, shown in Figure~\ref{fig:abs}.

 The cerium-doped samples exhibit the expected increase in UV absorption. More than that, by subtracting the host spectrum from a CgCAF12Ce0.10 spectrum, the band around 330~nm emerges (inset in figure~\ref{fig:abs}). We interpret that as the ground state absorption 4f$_{5/2}$ $\rightarrow$ 5d$_{1,2}$ in Ce$^{3+}$ ions~\cite{Wantana2018,Zuo2016,Chewpraditkul2012}. The absorbance, though, has increased beyond the range of this transition, and it is likely that Ce$^{4+}$ is contributing to it~\cite{Milewska2025}; however, we were unable to precisely identify this band.

 As we have demonstrated in previous works~\cite{Belancon2023,Sandrini2023,Belancon2021}, a cover material for silicon solar applications should exhibit UV absorption similar to that observed in the CgCAF12Ce0.10 and CgCAF12Ce0.25 samples, which is to absorb in the UV range without compromising the visible transmittance.

\begin{figure}[htp]
    \centering
    \includegraphics[width=1\linewidth]{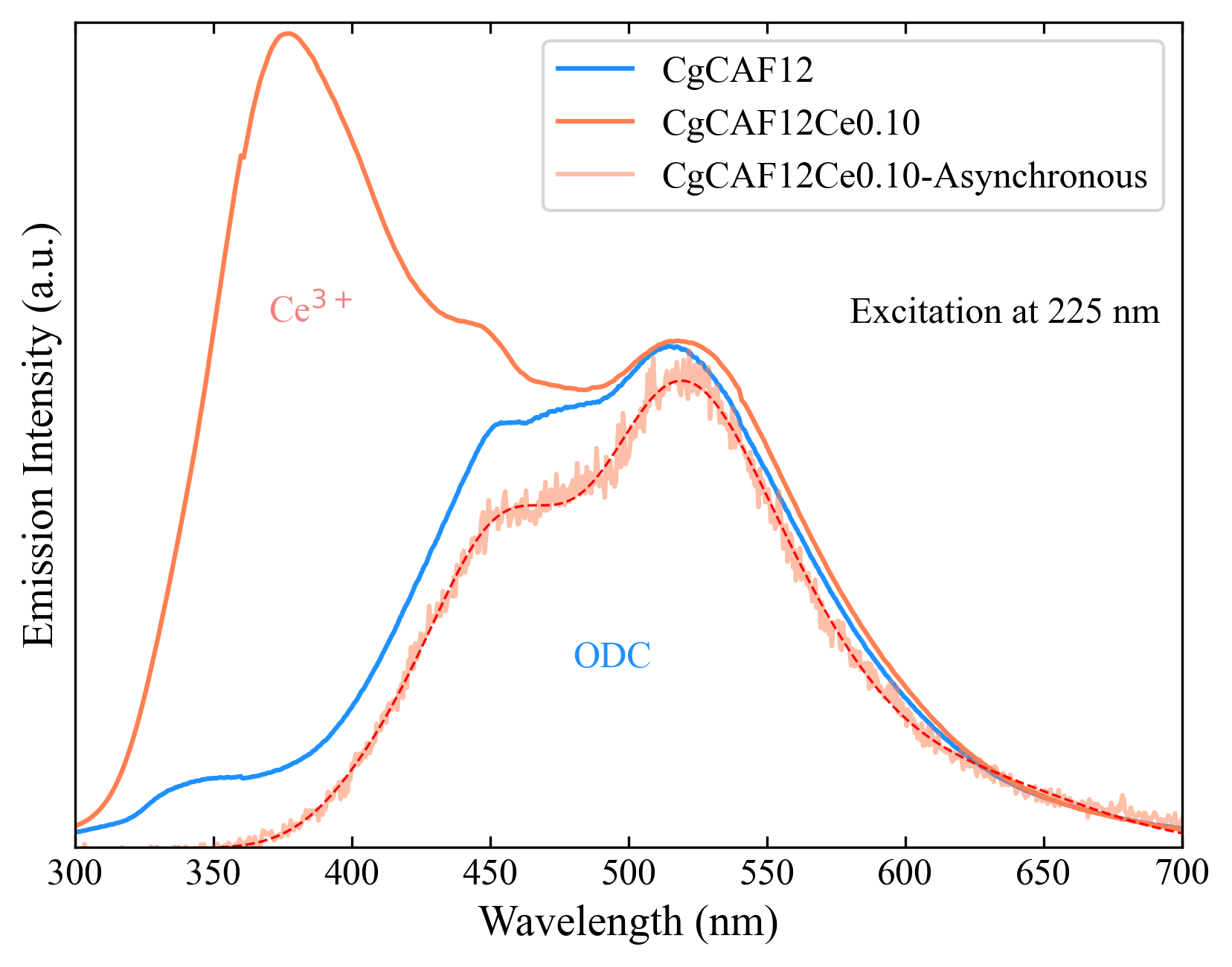}
    \caption{Emission spectra of matrix and a doped sample, including an asynchronous emission spectrum (smoothed in the red dashed curve) where fast components can not be detected. Intensities were adjusted for better data visualization.}
    \label{fig:em225}
\end{figure}

Next, the luminescence was investigated under various emission and excitation configurations. In Figure~\ref{fig:em225} we show some spectra selected to provide a precise insight into the luminescence properties.
First, by exciting at 225~nm, multiple broadbands from 400 to 600~nm (blue curve) were detected. This is the spectrum resulting in the pale green color under UV excitation observed in figure~\ref{fig:placeholder}, which is a feature of the CgCAF12 matrix. A similar spectrum had been reported in the literature~\cite{Trukhin2019}, and we interpreted that as an oxygen-deficiency center (ODC). More precisely, this emission band is the triplet luminescence (T$_1\xrightarrow{}$S$_0$)~\cite{Skuja1998,Salh2011,Garmysheva2022}. 

Second, the doped samples contain an additional emission band around 380~nm, which we attributed to the 5d-4f decay of Ce$^{3+}$. This band partially overlaps the ODC band, and, depending on the excitation wavelength or the cerium oxide concentration, it becomes far more intense than the ODC band, explaining why the pale green luminescence is not observed in the doped samples under UV illumination (Figure~\ref{fig:placeholder}).

The luminescence originating from an ODC triplet state exhibits a characteristic millisecond lifetime, opposed to the typically nanosecond lifetime from the ODC singlet-singlet transition (S$_1\xrightarrow{}$S$_0$) or Ce$^{3+}$ decays (5d$_1\xrightarrow{}$4f$_{5/2,7/2}$). To confirm this, we performed a second luminescence measurement in asynchronous mode using the internal chopper in the JP-8550 fluorometer. In brief, this configuration chops excitation light from the lamp and collects the emission from the sample after a time delay. We set this delay at 10 miliseconds, in such a way that only long lifetime components can be detected. The spectra for all samples is shown in Figure S1, while Figure~\ref {fig:em225} shows the result for the CgCAF12Ce0.10 sample. 

Once this configuration reduces the integrated signal reaching the detector, the spectrum becomes noisier; however, as shown, the main finding is that the faster 380~nm component has completely disappeared. The remaining spectrum matches that detected in the CgCAF12 matrix, confirming the hypothesis discussed above regarding the origin of the luminescence bands. 

Increasing the dopant concentration decreases the luminescence intensity from both ODC and Ce$^{3+}$ (also visually observed in figure~\ref{fig:placeholder}). Higher cerium oxide concentration can increase Ce$^{4+}$ proportion~\cite{Lertloypanyachai2026}, so our interpretation is that the higher optical absorption due to Ce$^{4+}$ reduces the excitation of both the ODC and Ce$^{3+}$ ions, while the 4f-5d resonant characteristic also induces self-quenching of the latter. Finally, the small bumps around $\sim$350~nm and $\sim$450~nm in the CgCAF12 spectrum are also much faster than the triplet decay, but this present study is not sufficient to fully identify them. 

\begin{figure}[htp]
    \centering
    \includegraphics[width=1\linewidth]{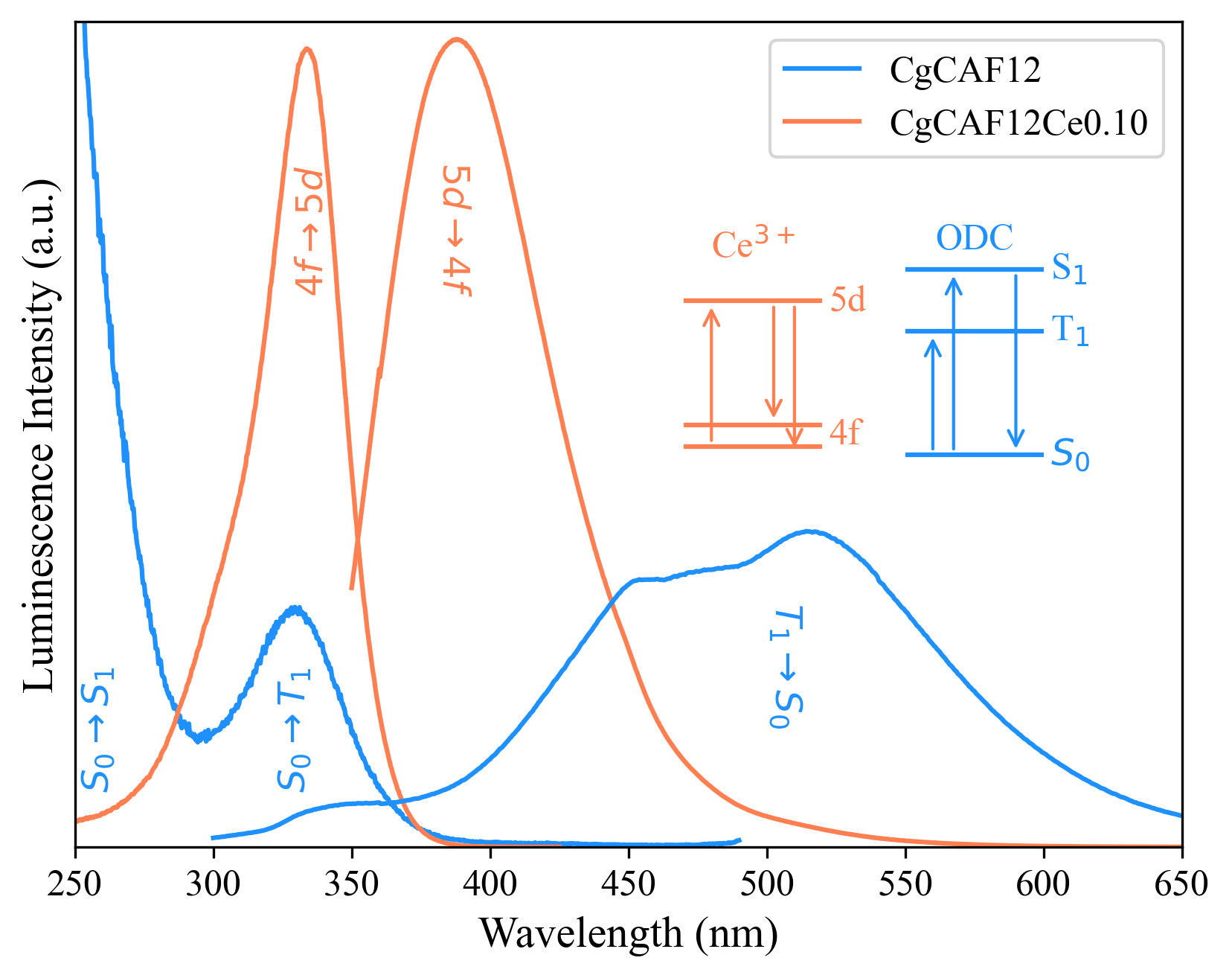}
    \caption{ODC (Ce$^{3+}$) emission under 225 (330)~nm pumping, and ODC excitation for 510 (440)~nm emission. Intensities were adjusted for better data presentation. A representative energy diagram for both species is also shown.}
    \label{fig:Ce}
\end{figure}

An optical excitation study was performed to identify the absorption spectrum of the emitting species in the UV range. As we mentioned earlier, different types of defects~\cite{Griscom2011,Trukhin2019,Skuja1998,Zatsepin2010} and both valence states of cerium are expected to absorb light in this range. In Figure~\ref {fig:Ce}, we can see the excitation spectra for both CgCAF12 and CgCAF12Ce0.10 samples, related to the ODC band at 510 nm and the Ce$^{3+}$ band at 440 nm. Although the latter band has a peak at 380 nm, we chose this wavelength because it enabled us to collect the entire excitation spectrum. 

In this way, the electronic transitions S$_0\rightarrow$S$_1$ ($\sim$250~nm) and S$_0 \rightarrow$T$_1$ ($\sim$330~nm) in ODC, as well as the 4f$\rightarrow$5d ($\sim$330~nm) in Ce$^{3+}$ were identified. This latter one matches exactly the absorption band demonstrated in the inset in figure~\ref{fig:abs}. It is important to note the overlap between the two excitation bands at 330~nm, however, we highlight that the intensities here were adjusted for better data presentation, and the emission spectra of CgCAF12Ce0.10 under 330~nm excitation exhibit only Ce$^{3+}$ 5d-4f decay. Indeed, we can fit this emission spectrum pretty well with two Gaussians, separated by $\sim$2000~cm$^{-1}$, which is the expected energy gap between the 4f$_{5/2}$ and 4f$_{7/2}$ levels.

Using the spectroscopic data shown above, we constructed a representative energy diagram, which is shown as an inset in figure~\ref{fig:Ce}. As we previously discussed, a variety of species contribute to the UV absorption in our samples, and further studies are necessary to provide a comprehensive description of the radiative process. For example, figure~\ref{fig:em225} shows Ce$^{3+}$ emission under 225~nm excitation; however, at this point, we are not able to confirm if we are exciting this ion directly, through its excitation tail below 250~nm; or if we are exciting the ODC $S_1$ state from where the 5d state in Ce$^{3+}$ gets populated. 

\begin{figure}[htp]
    \centering
    \includegraphics[width=1\linewidth]{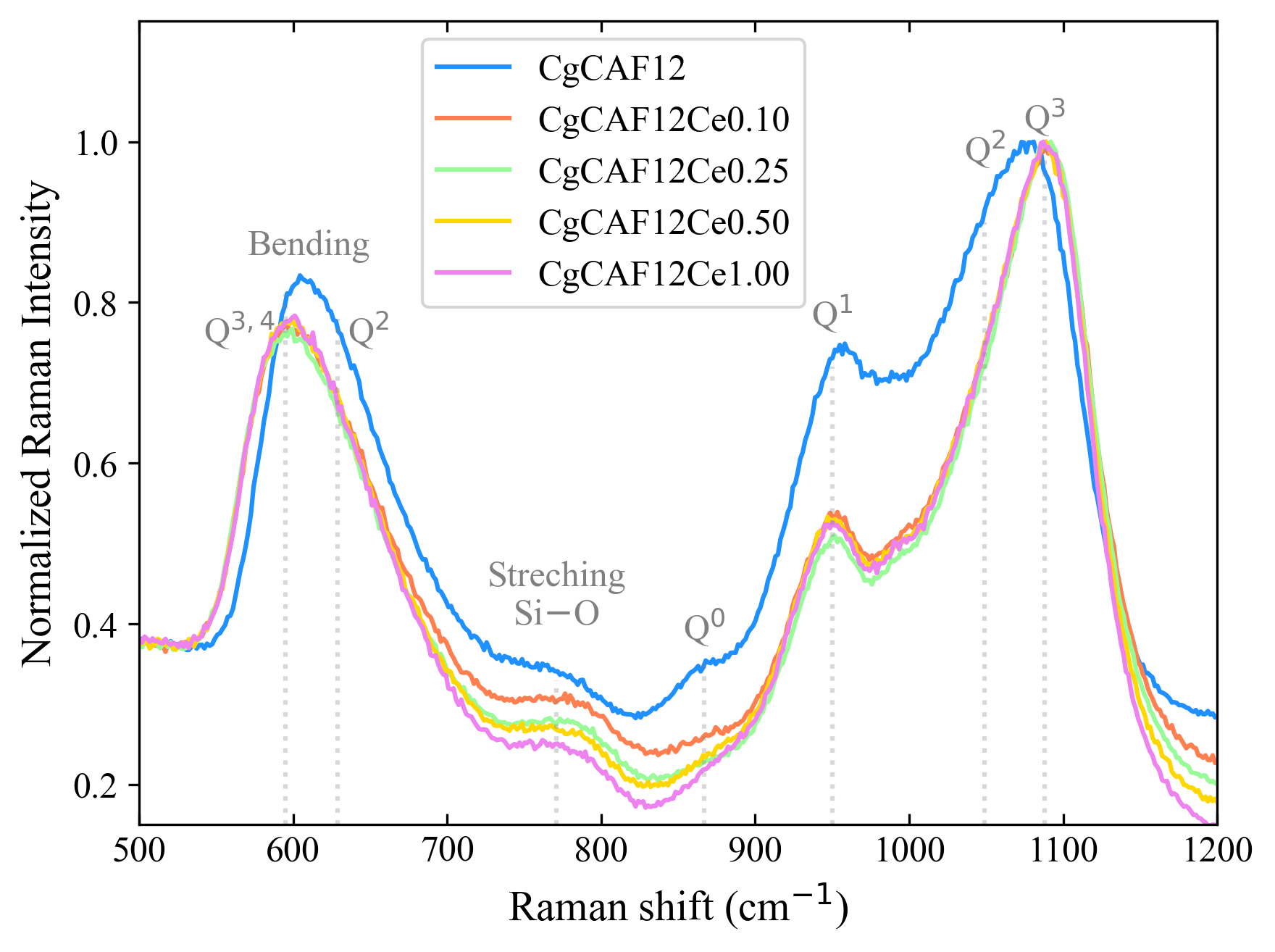}
    \caption{Raman spectra under 514~nm wavelength. The gray dotted lines indicate the positions of the main bands, as identified in the figure.}
    \label{fig:raman}
\end{figure}

Adding cerium oxide to silicates may also affect the glass network. Raman spectroscopy was used to investigate that, and the results are shown in Figure \ref{fig:raman}. 
Cerium significantly affected the Raman shifts, even at the lowest dopant concentration. The SiO$_2$ network is composed of units with different numbers of bridging oxygens (0-4), commonly referred to as Q$^n$ units, which give rise to Raman shifts around the positions indicated in the figure. Cerium doping decreases the Q$^{0,1}$ components, while it increases the Q$^{3}$ contribution near 1100~cm$^{-1}$. In the Si$-$O bending range ($\sim$600~cm$^{-1}$), the components Q$^{3,4}$ are also favored in relation to the Q$^{2}$. This indicates that cerium favors a more polymerized network~\cite{Qi2021}, as units with more bridged oxygens contribute proportionately more. We have demonstrated that increasing the CaF$_2$ content in CgCAF hosts depolymerizes the glass network~\cite{Valcarenghi2025}, an effect that is partially mitigated by cerium doping.

\begin{figure}[htp]
    \centering
    \includegraphics[width=1\linewidth]{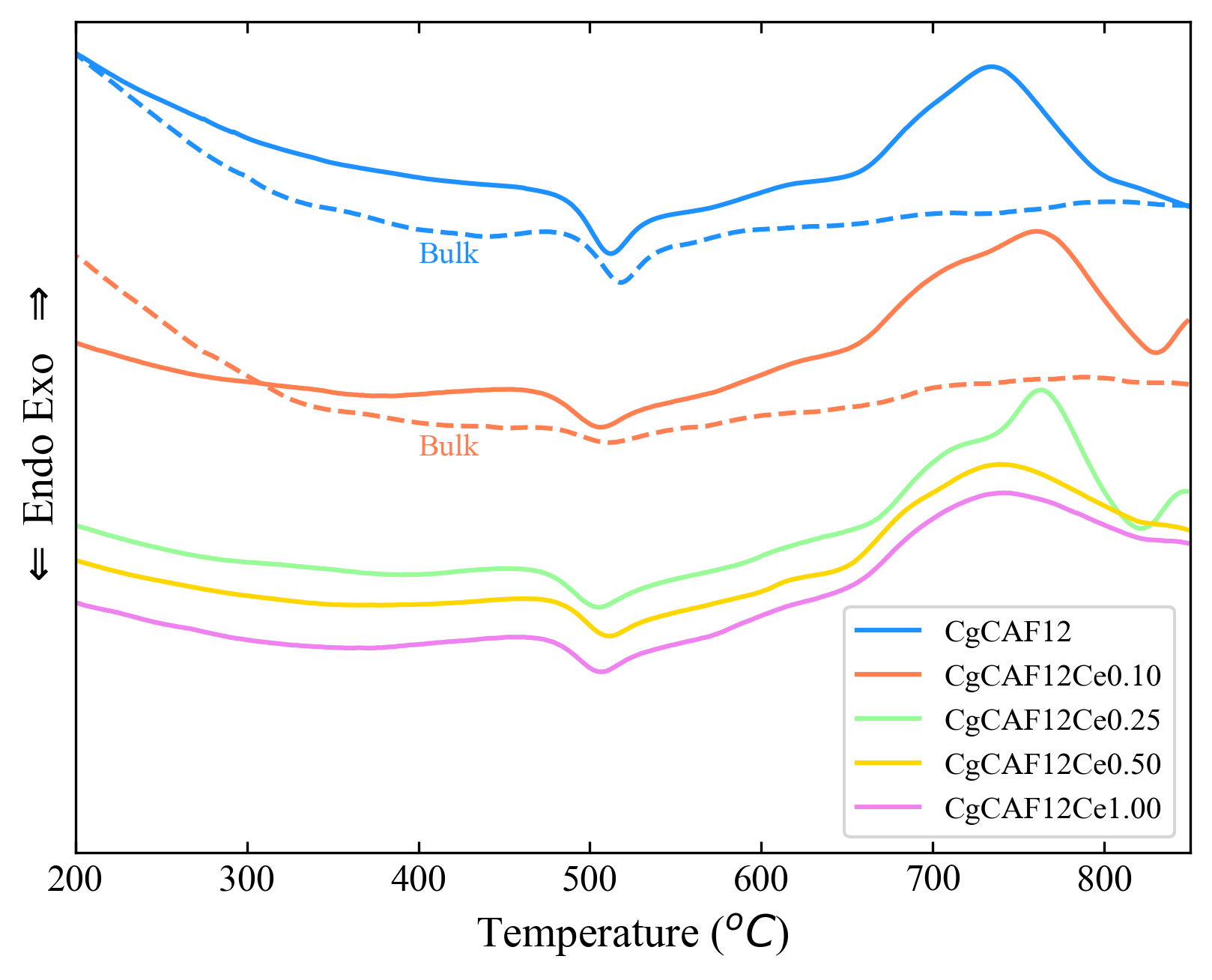}
    \caption{Differential thermal analysis from powder (continuous lines) and bulk (dashed lines) samples.}
    \label{fig:dsc}
\end{figure}

DSC measurements were performed in all samples (powdered) to see if cerium could affect the thermal properties of CgCAF12 matrix~\cite{Valcarenghi2025}. Bulk samples of the latter and the CgCAF12Ce0.10 were also submitted to this study. Figure~\ref{fig:dsc} shows the DSC curves obtained, and the characteristic temperatures are summarized in Table~\ref {tab:tg}.

\begin{table*}[htp]
\centering
\begin{tabular}{c|c|c|c|c|c}
\hline
\textbf{Sample} & \textbf{T$_g$ ($^\circ$C)} & \textbf{T$_x$ ($^\circ$C)} & \textbf{T$_p(1)$ ($^\circ$C)} & \textbf{T$_p(2)$ ($^\circ$C)} & \textbf{T$_x$-T$_g$ ($^\circ$C)}\\ \hline \hline
CGCAF12~\cite{Valcarenghi2025} & 501 & 656 & 735 & - & 155\\ \hline
CgCAF12Ce0.10 & 492 & 653 & 715 & 760 & 161\\ \hline
CgCAF12Ce0.25 & 493 & 663 & 718 & 763 & 170\\ \hline
CgCAF12Ce0.50 & 500 & 651 & 733 & - & 151\\ \hline
CgCAF12Ce1.00 & 492 & 653 & 733 & - & 161\\ \hline
\end{tabular}
\caption{Thermal parameters obtained from DSC measurements.}
\label{tab:tg}
\end{table*}
First of all, crystallization events above 650~\degree C were detected in all powdered samples, and they are not present in the curves for the bulk samples. This indicates that crystallization is concentrated at the surface of the CgCAF12 samples; meanwhile, between 650 and 800~\degree C, cerium incorporation alters the shape of the curve, suggesting that crystallization dynamics are changing. As one of our goals is to explore these samples to obtain glass-ceramics, it is fundamental to investigate these exothermic events further.

Based on the DSC results shown above, we decided to perform an in situ XRPD study from room temperature to 800~\degree C. The full experimental data for each sample contains the RT measurement before and after the experiment, and 14 patterns collected between 500-800~\degree C. In order to better present and discuss this data, only some selected spectra are shown next, while the entire set is available as supplementary material in figures S2-S6.

\begin{figure}[htp]
    \centering
    \includegraphics[width=1\linewidth]
    {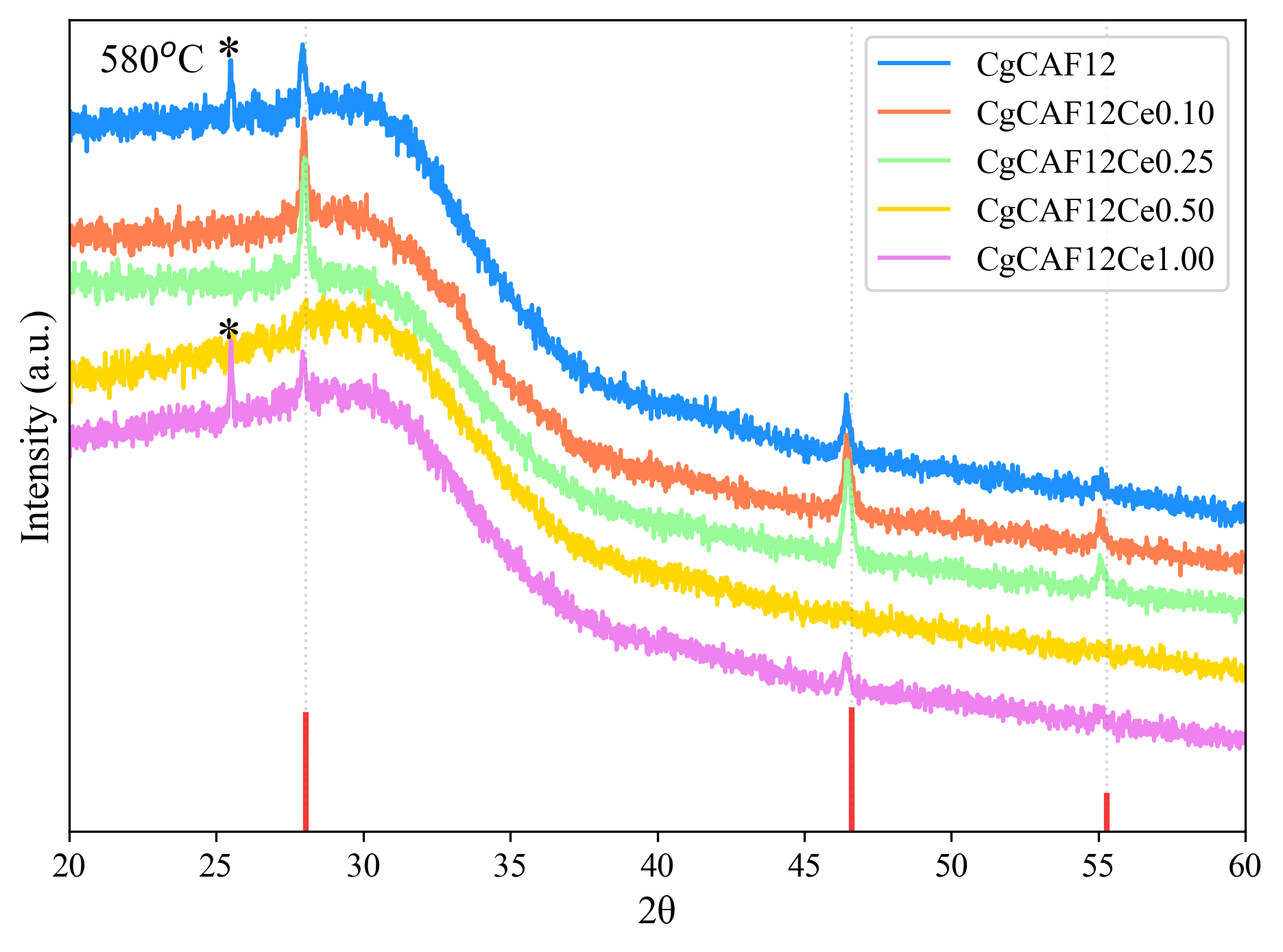}
    \caption{Crystallization observed between 500-600~\degree C. Red bars denote the main lines of CaF$_2$ crystals (ICDD PDF 04-021-7904) and the * marks the peak from the sample's holder (Al$_2$O$_3$).}
    \label{fig:xrd580}
\end{figure}

At 500~\degree C, some peaks from fluorite crystals (CaF$_2$, ICDD PDF 04-021-7904) started to appear in all samples except for the CgCAF12Ce0.50. In figure \ref{fig:xrd580} we show the patterns obtained at 580~\degree~C where this phase can be clearly seen.
For the first and last curves, a peak at 25.5~\degree from the sample holder (Al$_2$O$_3$) was also detected. As shown, CgCAF12Ce0.50 differs from the others in relation to fluorite crystallization. We highlight that all samples were subjected to the same procedures, yet the DSC data in table~\ref{tab:tg} show that the samples differ slightly in T$_g$. It is possible that during annealing at 480~\degree, fluorite nucleation was less effective in the CgCAF12Ce0.50 sample than in the others, but further studies are needed to elucidate this.

\begin{figure}[htp]
    \centering
    \includegraphics[width=1\linewidth]
    {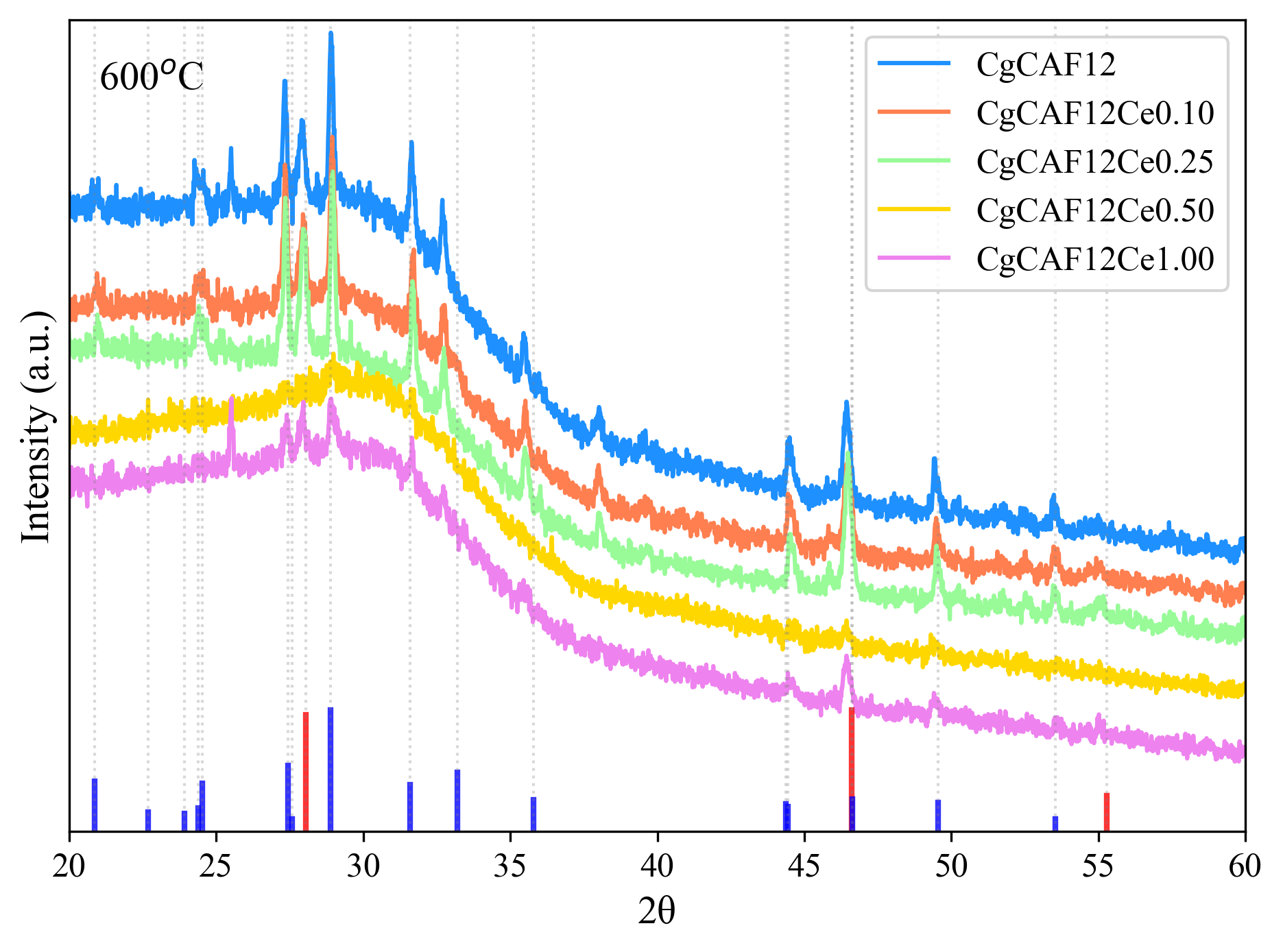}
    \caption{In situ PXRD at 600~\degree C evidencing xonotlite formation (blue bars, ICDD PDF 01-074-7586).}
    \label{fig:xrd600}
\end{figure}

Continuing the heating, several other peaks started to appear at 600~\degree C, and most of them are attributed to the xonotlite (Ca$_6$Si$_6$O$_{18}$H$_2$O, ICDD PDF 01-074-7586) as demonstrated in figure~\ref{fig:xrd600}.
This xonotlite pattern also matches the previously reported unknown phase on the CgCAF12 matrix~\cite{Valcarenghi2025}. Being a hydrate, the xonotlite phase is often obtained in calcium silicate hydrate mixes at much lower temperatures ($<$250~\degree C)~\cite{Yang2026,Mingione2025}. Upon heating, this phase is relatively stable until $\sim$800~\degree C, when it is expected to be converted into wollastonite (CaSiO$_3$)~\cite{Ogur2021}.

Some works on sodium calcium silicates rich in CaF$_2$ have also reported the presence of xonotlite at temperatures between 700-1000~\degree C~\cite{Miller2004,Hamzawy2014,Saito2024}. Our main hypothesis is that CaF$_2$ is likely to react with H$_2$O/OH$^-$ impurities that may come from the raw material, or from the atmosphere during glass melting, enabling crystallization of xonotlite, resulting also in F$^-$ modified xonotlite~\cite{dePablos-Martn2021,Kirste2024}.

\begin{figure}[htp]
    \centering
    \includegraphics[width=1\linewidth]
    {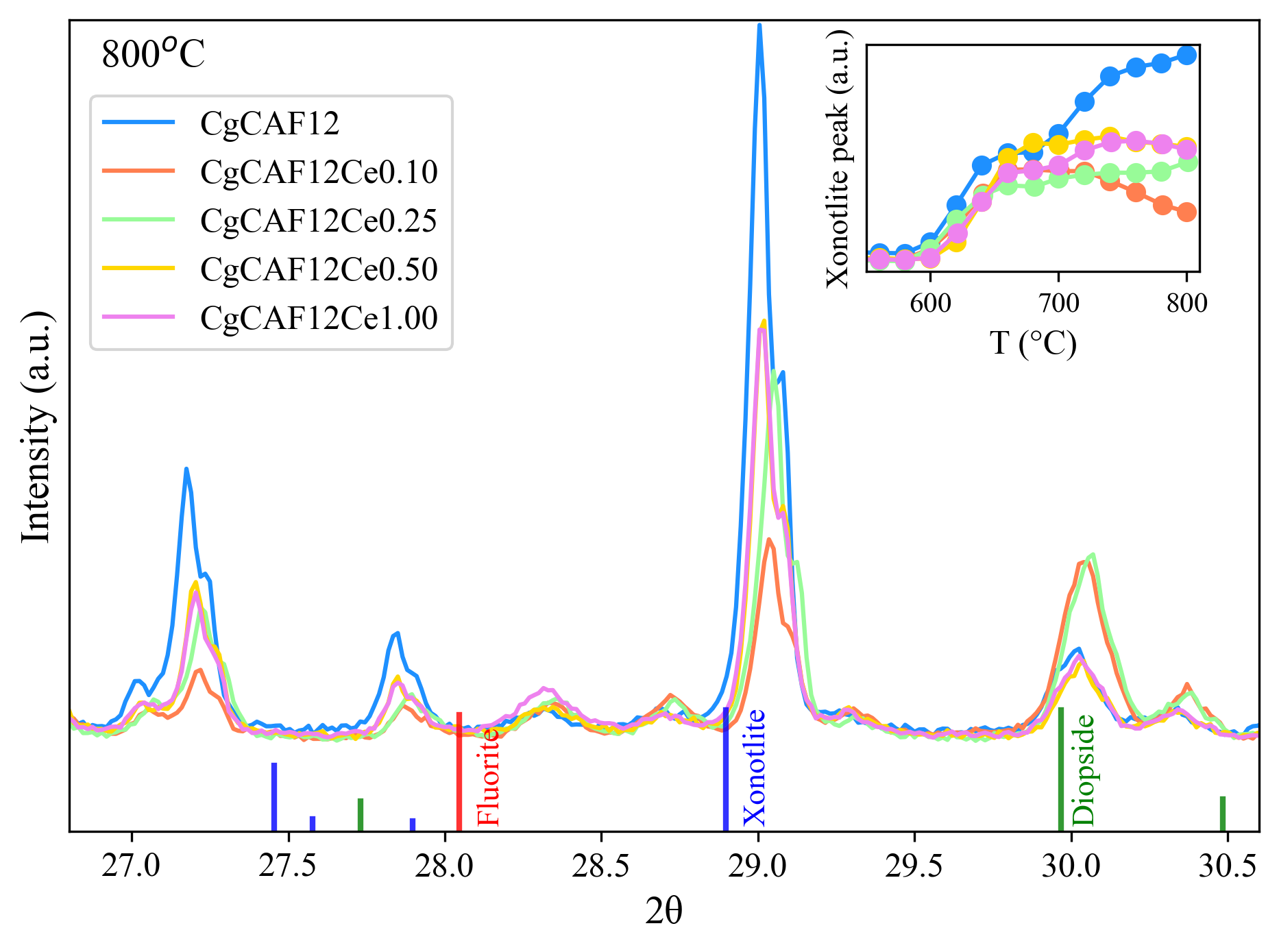}
    \caption{In situ PXRD diffraction patterns at 800~\degree C, evidencing the cerium effect in favoring xonotlite phase. The inset shows the intensity of the 29~\degree peak as a function of temperature for each sample.}
    \label{fig:xrd800}
\end{figure}

Beyond fluorite and xonotlite, other phases such as combeite (Na$_4$Ca$_4$Si$_6$O$_{18}$, ICDD PDF 04-008-0810) and diopside (CaMgSi$_2$O$_6$, ICDD PDF 04-013-1854) were detected above 600~\degree C. In figure \ref{fig:xrd800} we show a narrow section of the patterns collected at 800~\degree C for each sample, highlighting the most important features we have observed.
First of all, though there are still some peaks matching the fluorite pattern at 800~\degree C, they are less intense and are likely to be overlapped by peaks from another crystal. For example, the diopside is expected to have minor peaks near 28~\degree, as indicated in the figure.

The presence of other phases in small quantities and simultaneous growth with xonotlite is not discarded. For example, sodium silicate (Na$_2$Si$_2$O$_5$, ICDD PDF 04-014-8492), has main peaks at 15~\degree, 22~\degree, and $\sim$30~\degree, which are detected above 640~\degree C, though their proportions or linked evolution over temperature cannot confirm its presence. In all samples, combeite becomes the dominant phase at higher temperatures, as we previously observed~\cite {Valcarenghi2025}.

The inset in figure \ref{fig:xrd800} evidences the behavior of the main xonotlite peaks. We confirmed the same trend for at least 8 other xonotlite peaks. This phase forms around 600~\degree C in all samples, though the dynamics at higher temperatures differ. For the undoped sample, its intensity continued to increase in the range 700-800~\degree C. 

As shown in the supplementary materials S2-S6, between 640 and 780~\degree C, several unidentified phases are observed around the main 29~\degree~peak. These reflections can be due to cuspidine (Ca$_4$Si$_2$O$_7$F$_2$, ICDD 00-011-0066) and calcium silicate (Ca$_3$SiO$_5$, ICDD 00-017-0445)~\cite{Valcarenghi2025} phases, for example. Especially, for the main diopside peak at 30~\degree, we observed additional peaks at 15~\degree~and 49~\degree~that exhibit the same temperature evolution. These peaks are all highlighted in figures S2-S6, and further studies are required to fully identify these phases.

\begin{figure}[htp]
    \centering
    \includegraphics[width=1\linewidth]
    {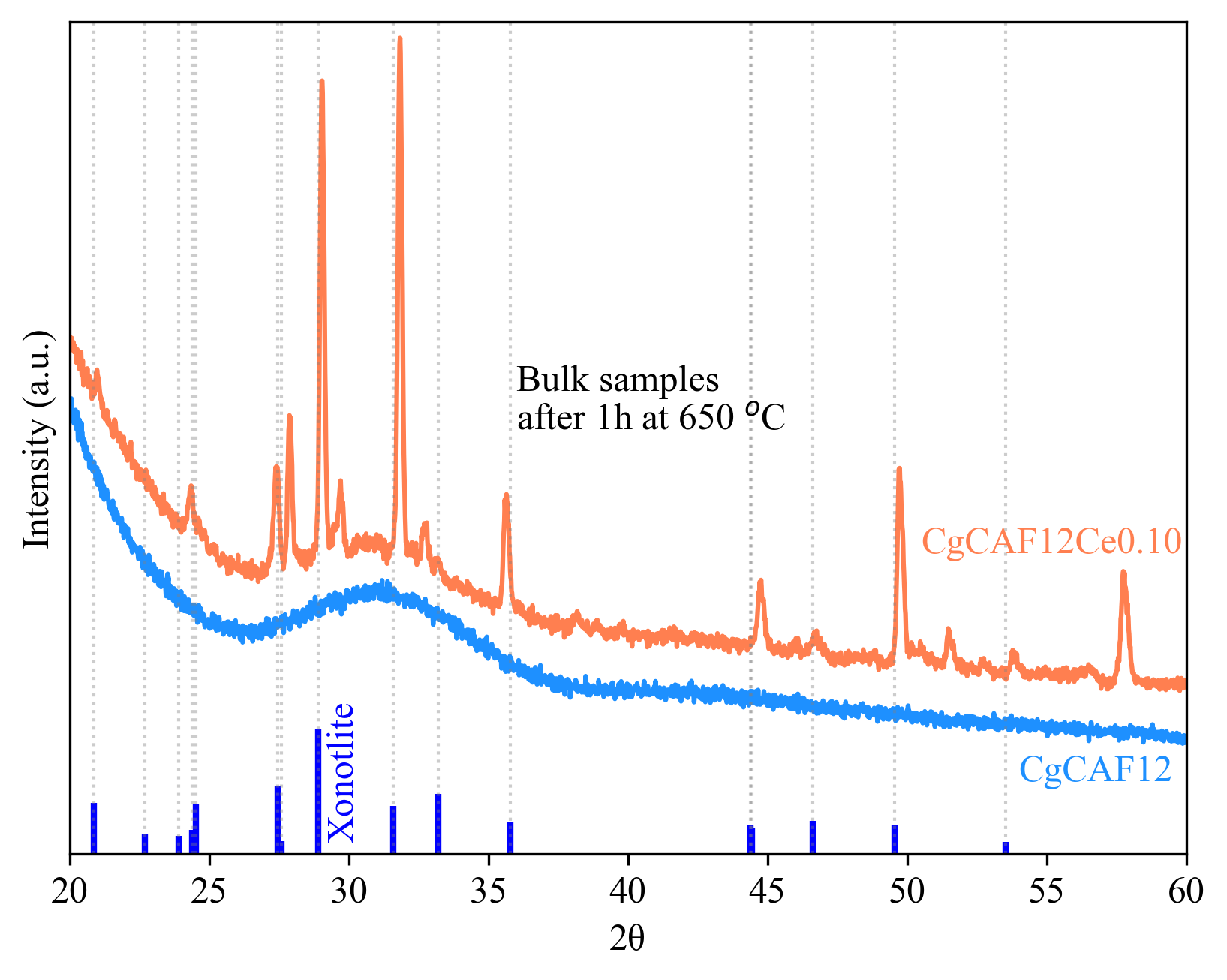}
    \caption{Room temperature XRD patterns of bulk samples annealed at 650~\degree C for 1 hour under air atmosphere.}
    \label{fig:pxrd_bulk}
\end{figure}

To conclude this investigation, CgCAF12 and CgCAF12Ce0.1 bulk samples were submitted to a last experiment. A furnace was preheated to 650~\degree C, and a platinum crucible containing discs from both samples was simultaneously placed inside it for 1 hour. After this annealing, the samples were removed from the furnace, cooled to room temperature, and subjected to an XRD experiment. The results are shown in figure~\ref{fig:pxrd_bulk}.

As shown, while the CgCAF12 surface remains a glassy phase, the CgCAF12Ce0.10 exhibits a diffraction pattern matching that of xonotlite. The relative intensities of some peaks have changed, though, and at least one peak is completely new to this study (at 57~\degree). We conclude that this phase forms on the sample's surface, but the mechanism underlying its formation, as well as the phase modification by F$^-$, remains to be confirmed. 

The ensemble of experimental data shown here demonstrates that xonotlite is forming on the surfaces of all samples. However, DSC indicates that crystallization events shift to lower temperatures in cerium oxide-doped samples, and the annealing at 650~\degree C confirmed this feature. The in situ PXRD, however, shows that xonotlite remains an important phase in the CgCAF12 sample at higher temperatures, and as we have demonstrated that cerium presence favors a more polymerized glass network, this feature can be linked to the xonotlite crystallization. 

In summary, this study provides fundamental information on the potential of the CgCAF12 oxyfluoride glass. First of all, its remarkable transparency~\cite{Valcarenghi2025} makes it interesting for optical applications. While Ce-doping successfully provided UV protection, the blue luminescence in the CgCAF12Ce0.10 sample indicates the material could be explored as a spectral converter. By aiming for this application, a future study should focus on optimizing the Ce$^{3+}$ proportion and luminescence. Once the CgCAF12Ce0.10 samples exhibited fluorite crystallization, controlling their growth could be a pathway to enhance luminescence properties~\cite{zhoulei2022}.

While fluorite crystals were detected just above 500~\degree C (except for the sample CgCAF12Ce0.50), the more significant phases detected above this temperature were xonotlite ($\sim$600~\degree C) and combeite ($\sim$700~\degree C). Additionally, cerium-doping favors an increase in the combeite/xonotlite ratio at higher temperatures.

Crystals may affect the homogeneity and viscosity of glass melts~\cite{Kilinc2024}, even at small proportions, and pose a challenge for processing large samples. Our work, however, shows that within the experimental detection limits of our PXRD study, there is no evidence of volume crystallization. On the surface, there is a range of at least $\sim$100~\degree C above T$_g$ where both the matrix and the cerium-doped  CgCAF12 remain quite stable against crystallization.

Beyond optical applications, the controlled crystallization of these phases can enable the transformation of the CgCAF12 glass into glass-ceramics with tailored thermal, mechanical~\cite{Levinskas2018,Liu2024,Saito2024}, and bioactive properties~\cite{Caland2024}. Specifically, for the xonotlite phase, an important goal is to understand the reactions that drive its nucleation and growth.

\section{Conclusion}

In summary, this work has demonstrated that cerium-doped CgCAF12 glasses can be obtained at a relatively low temperature (1200~\degree C) using about 75~wt\% of cullet from solar glass. The obtained samples have shown no signal of crystallization before $\sim($T$_g$+100)~\degree C is reached, indicating significant stability against crystallization. Regarding spectroscopic properties, all samples showed high transparency, and cerium doping was effective in controlling UV transmission, resulting in blue emission from Ce$^{3+}$ ions.

Production of solar glass is significant, and to reduce environmental impact, it is mandatory to find pathways for its reuse and recycling. The CgCAF12 samples investigated in this work have significant potential to address this challenge. These glasses exhibit high transparency from UV to IR, and cerium oxide addition was effective in providing UV absorption. The dopant also affected the glass network, which became more polymerized. The DSC and the in situ high-temperature PXRD successfully demonstrated the crystallization of fluorite, xonotlite, and combeite phases, as well as minor proportions of diopside and other phases that will require further studies to be fully ascribed. In addition, developing techniques to process this material and control its crystallization are required to fully realize its potential applications.

\section{Acknowledgement}

The authors acknowledge the financial support from CNPq (grants 409475/2021-1, 402473/2023-0, and 304060/2023-2) and from Centre – Val de Loire Region et interregional Loire FEDER-FSE+ 2021-2027 Program Pyrometer Convention n\degree 2025-00047653.

This work was also supported by Le Studium, Loire Valley Institute for Advanced Studies, under the Smart Loire Valley Program.

\bibliographystyle{elsarticle-num} 
\bibliography{references}

\end{document}